\documentclass[prl,showpacs,floatfix,twocolumn]{revtex4}

\usepackage[paperwidth=8.5in,paperheight=11in,centering,hmargin=0.75in,vmargin=1in]{geometry}
\usepackage{xfrac}

\usepackage{array}
\usepackage{multirow}
\usepackage{color}

\usepackage{booktabs}
\usepackage{graphicx}

\usepackage{tabularx}
\newcolumntype{Y}{>{\centering\arraybackslash}X}
\usepackage{epsfig,graphicx}

\newcommand{\presubspace}{-18pt}
\newcommand{\postsubspace}{-15pt}
\newcommand{\CoV}{NaCa$_2$Co$_2$V$_3$O$_{12}$}
\newcommand{\CoGe}{CaY$_2$Co$_2$Ge$_3$O$_{12}$}

\begin{document}

\title[Magnetodielectric Effects in Co-Garnets]
{Magnetodielectric effects at quantum critical fields in cobalt-containing garnets} 

\author{Abbey J. Neer} 
\affiliation{Department of Chemistry, University of Southern California, Los Angeles, CA 90089, USA.}

\author{Veronika A. Fischer}
\affiliation{Department of Chemistry, University of Southern California, Los Angeles, CA 90089, USA.}

\author{Michelle Zheng}
\affiliation{Department of Chemistry, University of Southern California, Los Angeles, CA 90089, USA.}

\author{Nicole R. Spence}
\affiliation{Department of Chemistry, University of Southern California, Los Angeles, CA 90089, USA.}

\author{Clayton Cozzan}
\affiliation{Materials Department, University California, Santa Barbara, CA 93106, USA.}


\author{Mingqiang Gu}
\affiliation{Department of Materials Science and Engineering, Northwestern University, Evanston, Illinois 60208, USA.}

\author{James M. Rondinelli}
\affiliation{Department of Materials Science and Engineering, Northwestern University, Evanston, Illinois 60208, USA.}

\author{Craig M. Brown} 
\affiliation{NIST Center for Neutron Research, Gaithersburg, Maryland 20899-8562, USA}

\author{Brent C. Melot$^{*}$} 
\affiliation{Department of Chemistry, University of Southern California, Los Angeles, CA 90089, USA.}
\email{melot@usc.edu}

\begin{abstract}
Here we present a comparative study of the magnetic and crystal chemical properties of two Co$^{2+}$ containing garnets, NaCa$_2$Co$_2$V$_3$O$_{12}$ and CaY$_2$Co$_2$Ge$_3$O$_{12}$.
Both phases exhibit the onset of antiferromagnetic order at 8\,K and 6\,K respectively, as well as field-induced transitions in their magnetization at 1\,T and around 11\,T. 
We find these field-dependent transitions correspond to quantum critical points that result in the suppression of antiferromagnetic order and that these transitions can be clearly seen using magnetocapacitance measurements. 
Finally, we perform detailed crystal-chemistry analyses and complimentary density functional theory calculations to show that changes in the local environment of the Co-ions are responsible for differences in the two magnetic structures and their respective properties. 

\end{abstract}

\maketitle
\section{Introduction}

Multiferroics, which simultaneously exhibit more than one ferroic property, offer unique opportunities to study the relationship between crystal structure and physical properties.
In particular, single phase materials that exhibit a cross-coupling between their magnetic and electrical polarizations can give rise to novel functionality like the ability to manipulate magnetizations through applied voltages or vice versa.\cite{Schmid1994}
%
Yet, as Spaldin (then Hill) explained, magnetism and ferroelectricity are  intrinsically contraindicated because of fundamental differences in the bonding required, making the design of new multiferroics a fascinating challenge.~\cite{Hill2000}

Several routes have been explored in recent years to overcome this materials chemistry challenge and enhance the strength of spin-lattice coupling achievable.
In the cubic spinel CoCr$_2$O$_4$, as well as LiCu$_2$O$_2$ and BiFeO$_3$, helical ordering of the magnetic moments breaks the spatial inversion symmetry.~\cite{Lawes2006,Yamasaki2006,Seki2008,Bush2012,Lee2013,Kimura2003}
Structural topologies with strong competition between nearest- and next-nearest neighbors have also been shown to create highly frustrated systems that can lead to  the evolution of electrical polarizations.
This is exemplified in TbMnO$_3$ where the reversal of the Ising moments on Tb$^{3+}$ induces a reorientation of the Mn$^{3+}$ spins and triggers an elongation along the $a$-axis.\cite{Kimura2003}
Similarly, commensurate-incommensurate magnetic transitions were reported to give rise to large anomalies in the magnetic field-dependent dielectric properties of orthorhombic manganites $R$Mn$_2$O$_5$ (where $R$=Tb, Dy, Ho).\cite{Hur2004,Kida2008}

Another challenge in the development of new magnetoelectric phases is that accurate measurements of electrical polarizations require mm-sized single crystals in order to detect the extremely small currents that are typically found.
Crystal growth is rarely straightforward so techniques capable of screening polycrystalline samples are highly desirable.~\cite{Sparks2014}
In response, magnetocapacitance measurements have quickly grown to be an essential tool since the temperature- or field-dependent capacitance of a well-sintered pellet can often reveal very minor changes in charge localization or expansions/contractions in the unit cell.~\cite{Lawes2003}
Magnetodielectric measurements can also be useful in identifying symmetry-breaking changes in the magnetic moments that would normally preclude the evolution of macroscopic polarizations based on the nuclear symmetry alone.~\cite{Lawes2009}

A few of us recently reported the observation of a quasi-one-dimensional magnetic order in the garnet CaY$_2$Co$_2$Ge$_3$O$_{12}$, where the Co$^{2+}$ ions on the octahedral site adopt an antiferromagnetic ground state with the moments fixed along the body diagonals of the unit cell.~\cite{Neer2017}
This highly anisotropic orientation of the spins forms discrete antiferromangetic rods that were found to undergo a magnetic-field driven quantum critical phase transition above fields of 6\,T.
This motivated us to look for analogous materials that could be used to discriminate whether the one-dimensional magnetic structure was necessary to realize quantum criticality, and ultimately led us to another Co-containing garnet, \CoV.
%
%
In both the germanate and vanadate phases, the lattice, magnetic cation, and magnetic topologies are effectively identical, providing an ideal comparison where only the diamagnetic portions of the host structure differ.
Herein, we report that NaCa$_2$Co$_2$V$_3$O$_{12}$ shows similar signatures of quantum criticality, albeit at substantially larger fields, despite exhibiting a radically different magnetic structure.
We also observe signatures of these quantum phase transitions in the field-dependent magnetocapacitance data, suggesting a correlation exists between the suppression of magnetic order and the magnetodielectric properties.
Taken together, these results demonstrate that Co-containing garnets offer an exciting class of materials for exploring fundamental nature of magnetic interactions in the solid state.



\vspace{\presubspace}
\section{Methods}

\vspace{-10pt}
\subsection{Materials Preparation}
\vspace{\postsubspace}
Polycrystalline powders of NaCa$_2$Co$_2$V$_3$O$_{12}$ were prepared by grinding stoichiometric ratios of CaCO$_3$, Na$_2$CO$_3$, Co(C$_2$O$_4$)$\cdot2$H$_2$O and V$_2$O$_5$ and pressing into pellets before firing in air through a multistage heat treatment. 
All pellets were isolated from the ZrO$_2$ crucible using a sacrificial layer of powder with the same stoichiometry as the target phase.
Co(C$_2$O$_4$)$\cdot2$H$_2$O was freshly prepared in house by precipitating a solution of Co(SO$_4$)$\cdot7$H$_2$O with an excess of oxalic acid and drying at room temperature overnight.
%
The first calcination was performed at 500$^{\circ}$C for 24 hours, at which point the pellets were ground and mixed well before re-pressing into pellets and heating at 850$^{\circ}$C in 24 hour increments until phase pure, usually requiring one to two treatments.
For final densification, powders were loaded into a 9\,mm carbon die and spark plasma sintered (SPS) at 600$^{\circ}$C for 5 minutes until 8\,kN of pressure was achieved to produce pellets with densities greater than 85\%. 

%

%

\vspace{\presubspace}
\subsection{Neutron Diffraction}
\vspace{\postsubspace}
Polycrystalline powders were loaded into vanadium cans and placed in a helium flow cryostat. 
Data sets were collected at 300\,K, 50\,K and 3\,K using the BT-1 high resolution neutron powder diffractometer at the NIST Center for Neutron Research. 
A Ge(311) monochromator with a 75$^{\circ}$ take-off angle, $\lambda$ = 2.0787(2)\AA\/\, and in-pile collimation of 60 minutes of arc were used. 
Data were collected over the range of 1.3$^{\circ}$-166.3$^{\circ}$ in scattering angle (2$\theta$) with a step size of 0.05$^{\circ}$. 
\vspace{\presubspace}
\subsection{Physical Property Measurements}
\vspace{\postsubspace}
Temperature and field dependent magnetic susceptibility as well as specific heat measurements were collected using a Quantum Design 14T Dynacool Physical Property Measurement System. 
For the specific heat measurements powders of NaCa$_2$Co$_2$V$_3$O$_{12}$ were ground together with equal parts silver in order to increase thermal coupling to the sample stage. 
The contribution of the silver was measured separately and subtracted.~\cite{Tari2003}

%
Magnetocapacitance measurements were performed on pellets that were densified using SPS with calculated densities between 86-98\%.
Dense pellets of the compound of interest were painted with Ag-epoxy (Epotek  EE129-4) and an epoxy-coated copper wire were attached to act as a parallel-plate capacitor.
The edges of the pellet were then sanded to ensure no short circuits are created between the sides of the electrode.
Shielded stainless steel co-axial cables were affixed to the electrode face and to the top of the custom-built measurement probe.
The capacitance was measured on a high precision capacitance bridge Andeen-Hagerling 2700A at a frequency of 1\,kHz interfaced to the same PPMS described above.
In order to rule out any effects due to magnetoresistance or leakage current through the pellet, AC Impedance measurements were performed over an array of frequencies at various fields.~\cite{Catalan2006}
As seen in SI Figure 1, neither material shows any evidence for field-dependent changes in the impedance, which is in agreement with the extremely low dielectric loss ($\tan{\delta} \approx$ 10$^{-4}$) obtained for the pellets at 2K.

\vspace{\presubspace}
\section{Density Functional Theory Calculations}
\vspace{\postsubspace}

Density functional theory (DFT) calculations were performed using Vienna Ab-initio Simulation Package (VASP).~\cite{Kresse1996a,Kresse1999}
The projector augmented wave (PAW) method ~\cite{Blochl1994a} is used to treat the core and valence electrons using the following electronic configurations: 2p$^6$3s$^1$ for Na, 2p$^6$2s$^2$ for Ca, 5s$^2$4s$^2$4p$^6$4d$^1$ for Y, 3s$^2$3p$^6$4s$^2$3d$^7$ for Co, 3d$^5$ for V, 4p$^4$ for Ge, 2s$^2$2p$^4$ for O.
The Brillouin zone is sampled using a $4\times4\times4$ $\Gamma$-centered Monkhorst-Pack $k$-point mesh and integrations are performed using Gaussian smearing with a width of 20\,meV.
Spin-orbit coupling (SOC) is considered and the revised Perdew-Burke-Ernzerhof (PBE) for solids (PBE-sol) \cite{Ropo2008} are used in our calculation. 
The electron correlation of the Co is considered within the spin-polarized DFT+$U$,~\cite{Dudarev1998} with an effective Hubbard correction of $U_{eff}$=2\,eV. 
Non-collinear magnetic spin configurations are taken into account to represent the magnetic ground state in these materials.

\begin{figure}[b]
\includegraphics[width=0.45\textwidth]{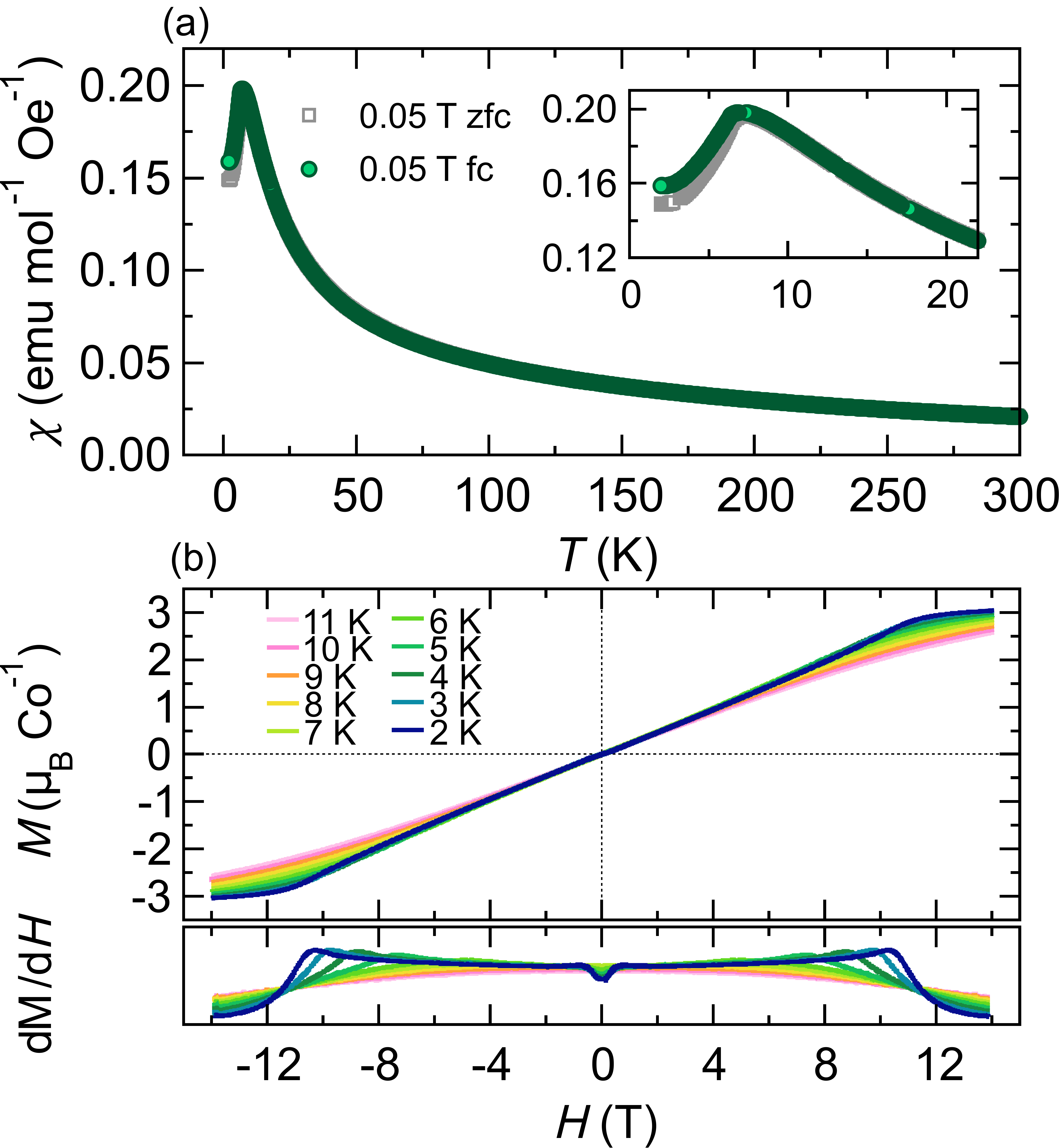}
\caption{(a.) Temperature dependent magnetic susceptibility of \CoV with an antiferromagnetic transition around 6K. (b.) Isothermal magnetization taken below and above the ordering temperature with the respected derivatives below. 1 Oe = (1000/$\pi$) A/m.}
\label{fig:magnetism}
\end{figure}

\vspace{-15pt}
\section{Results and Discussion}
\vspace{-5pt}

%
%
%
Figure \ref{fig:magnetism} (a) shows the temperature-dependent magnetic susceptibility of \CoV, which indicates a transition to an antiferromagnetic ground state around 6\,K.
Fitting the high-temperature region (200-300\,K) of the susceptibility to the Curie-Weiss equation and including a temperature independent paramagnetic contribution, $\chi$=C/(T-$\Theta_{CW}$)+$\chi_0$, yields a $\Theta_{CW}$ of -44\,K, an effective paramagnetic moment of 7.33$\mu_B$ per formula unit (5.18$\mu_B$ per Co) and $\chi_0$=6.6$\times 10^{-4}$ emu mol$^{-1}$ Oe$^{-1}$.
This moment is in close agreement with the expected value for Co$^{2+}$ in a high-spin octahedral coordination environment ($d^7$, $S$=3/2, $L$=3), when the orbital moment is unquenched and decoupled from that of the spin ($\mu_{L+S}$ =$\sqrt{4S(S+1)+L(L+1)}$).\cite{Day1960}
The negative sign of $\Theta_{CW}$ indicates that antiferromagnetic exchange is dominant between the spins and there is only a modest suppression of the ordering temperature ($\Theta/T_N$ = 5.5).

\begin{figure}[b]
\includegraphics[width=0.35\textwidth]{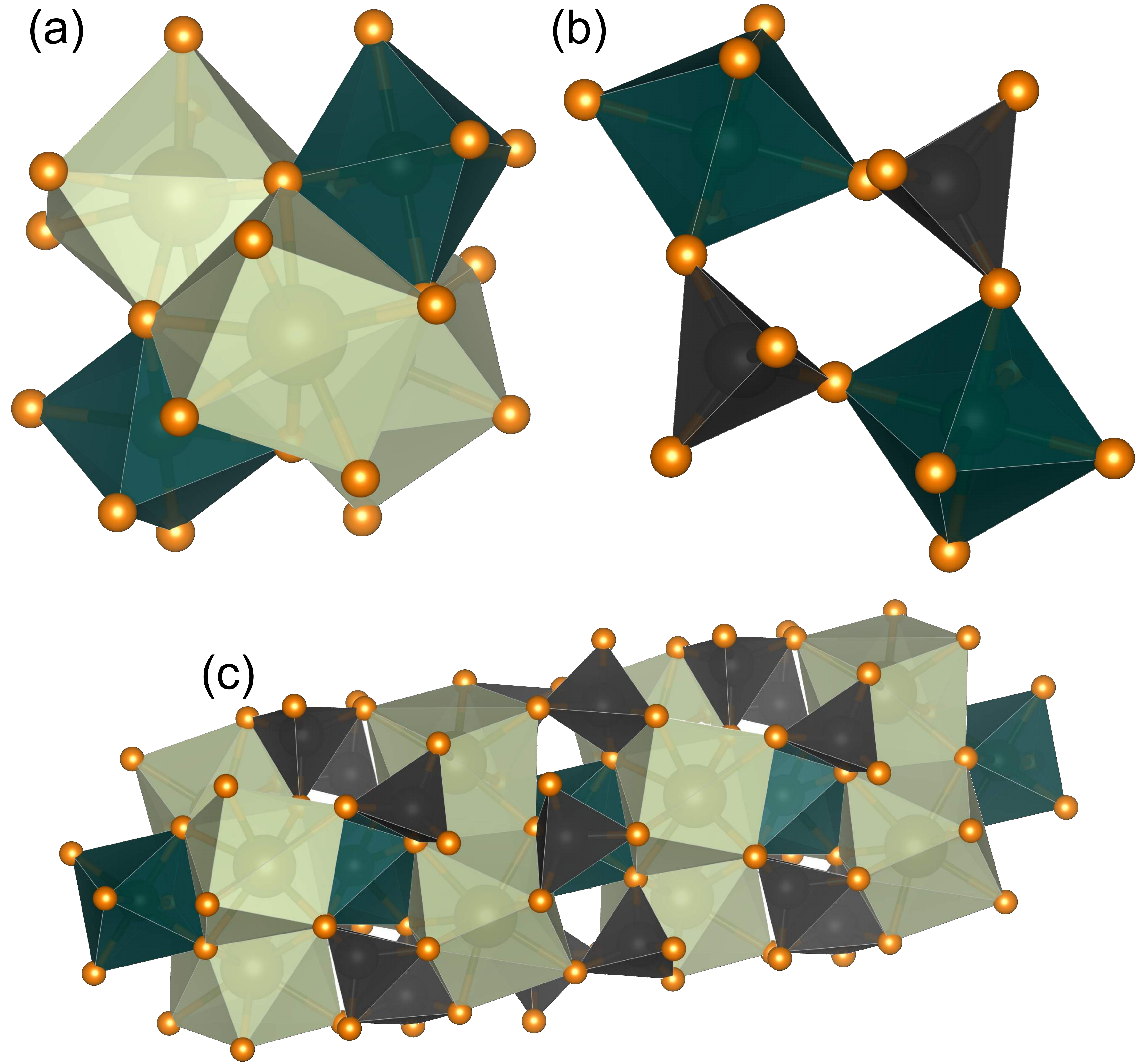}
\caption{ (a) Octahedral site connectivity through the cubic positions. (b.) Connectivity of the octahedra and tetrahedra. (c) Connectivity of a rod within the garnet structure. These rods consist of octahedrally coordinated ions bridged together through the edge of the cubic site \textit{via} along shared oxygen bonds. The are oriented along the body diagonal of the cubic unit cell. Oxygen is shown in orange, Co ions in green, Na/Ca in yellow, and V in gray.}
\label{fig:sublattice}
\end{figure}

The suppressed ordering temperature can be understood by considering that garnets consist of a network of $B$O$_6$ octahedra linked together at their corners by $A$O$_4$ tetrahedra and $R$O$_8$ dodecahedra on the edges. 
As O'Keefe first highlighted, and we later demonstrated in \CoGe, the octahedral sublattice forms rods, as illustrated in Figure \ref{fig:sublattice} (c), which run along the body diagonals of the unit cell.\cite{Andersson1977}
Given there is no direct connectivity between adjacent octahedral sites, more complex superexchange pathways must mediate the magnetic interactions either through the tetrahedral/cubic sites or along the edges of the polyhedra through super-superexchange pathways exclusively involving oxygen [see Figures \ref{fig:sublattice} (a) and (b)]. 
The presence of so many competing exchange pathways that must be simultaneously satisfied to reach the ground state magnetic structure is likely the cause of the suppressed ordering temperature.
%
%
%
%

Below the ordering temperature of 6\,K, the isothermal magnetization of  \CoV\/ shows a deviation from the linear response expected for an antiferromagnetically ordered system of spins at fields beginning around 10\,T [see Figure \ref{fig:magnetism}(b)].
A more faint deviation around 1\,T is seen in the $dM/dH$ curve of Figure \ref{fig:magnetism} (b).
Field-induced magnetic transitions are fairly common in divalent Co compounds because of its substantial anisotropic character.
Measurements were taken between 2\,K and 11\,K, to track the field-induced transition passed the ordering temperature, with the curves flattening out toward the traditional linear response for a paramagnetic state by 11K.

\begin{figure}[t]
	\includegraphics[width=0.4\textwidth]{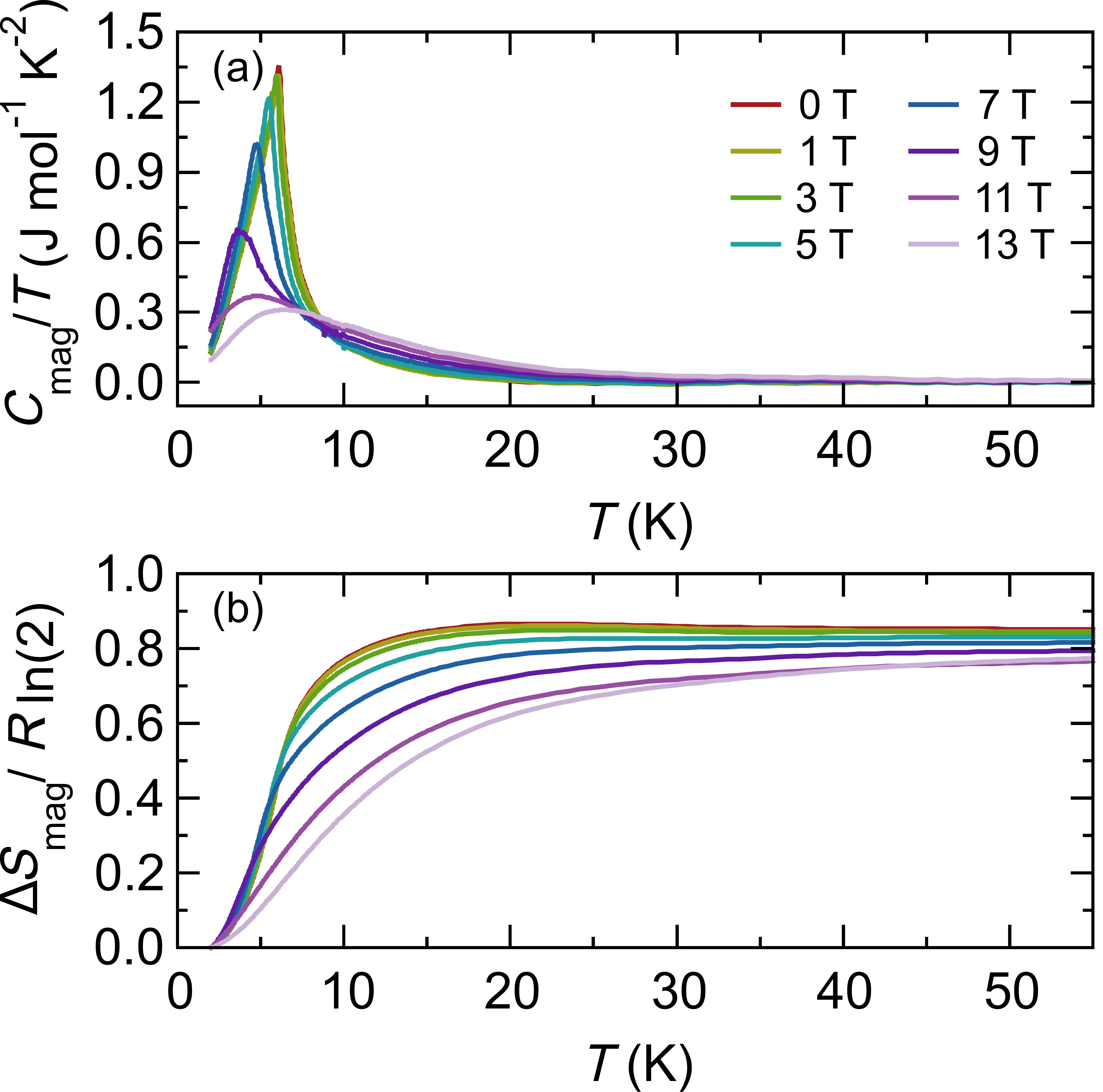}\caption{(a.) Field-dependent specific heat measurements. Note the Lambda anomaly corresponding to the three-dimensional ordering of NaCa$_2$Co$_2$V$_3$O$_{12}$ is gradually suppressed towards a broad anaomly more reminscent of a paramagnetic states above fields of 11T. (b.) Normalized entropy released around the magnetic transition normalized by the theoretical entropy expected for an S=$\sfrac{1}{2}$ Kramers ion like Co$^{2+}$.}\label{fig:heatcapacity}
\end{figure}

The field-dependence of the magnetization of \CoV\/ lead us to map out these transitions using specific heat measurements, shown in Figure \ref{fig:heatcapacity} (a).
In the absence of a magnetic field, a sharp lambda anomaly is seen around 6\,K, which agrees with the onset of three-dimensional order in the temperature-dependent susceptibility.
At small fields, between 0\,T and 1\,T, no significant change in the peak position is noticeable, but in fields starting at 9\,T the sharp anomaly is significantly blunted and smears out to resemble a broad feature reminiscent of a Schottky anomaly suggesting the system transitions into a field-polarized paramagnetic state. 
This is supported by the decreasing magnetic entropy with increasing field strength that is observed in Figure \ref{fig:heatcapacity} (b).

\begin{figure}[b]
	\includegraphics[width=0.4\textwidth]{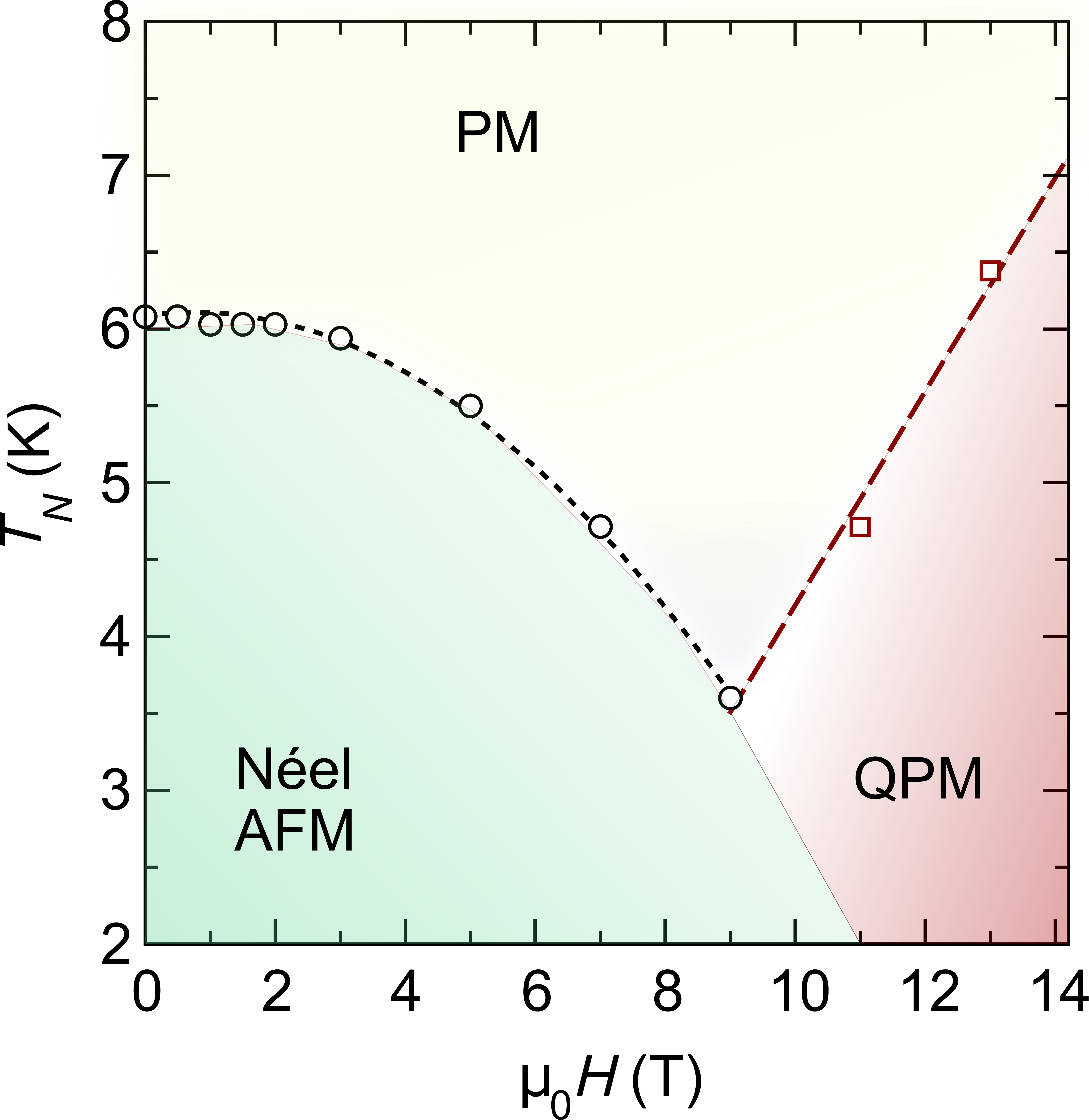}\caption{Magnetic Phase Diagram of \CoV\/ as constructed from the transition temperatures shown in Figure \ref{fig:heatcapacity} (a). PM = paramagnet, AFM = antiferromagnet, and QPM = quantum paramagnet, where spins are disorded through quantum fluctuations. A quantum critical point occurs near 11\,T in both the heat capacity and magnetization measurements.}\label{fig:phase}
\end{figure}

Combining the magnetization and specific heat measurements, a magnetic phase diagram was constructed from the observed transition temperatures as illustrated in Figure \ref{fig:phase}. 
Fitting a quadratic response to the field-dependence of the antiferromagnetic transition temperature suggests a critical magnetic field near 11\,T very similar to what was fond in \CoGe\/ albeit at much stronger fields.
%

\begin{figure}[t]
	\includegraphics[width=0.4\textwidth]{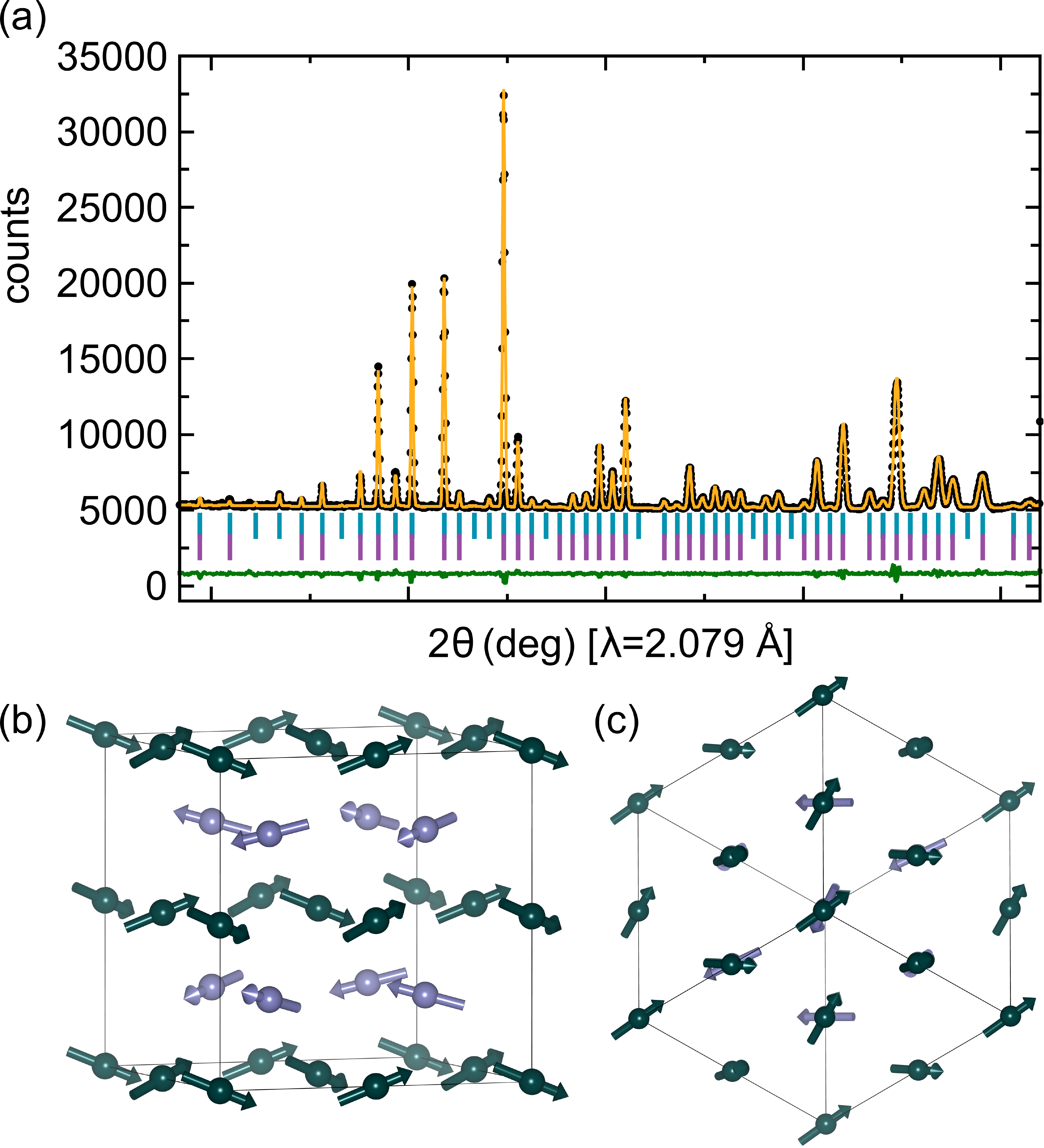}
	\caption{(a) Results of the Rietveld refinement of the nuclear and magnetic contributions to against the 3\,K powder neutron diffraction data, $R_{Bragg}^{nuc}$=1.5\%, $R_{mag}$=9.0\%. The data is shown as unfilled black circles whereas the solid orange and green lines show the fit and difference curves respectively. The top row of ticks represents the nuclear contribution while the bottom show peaks with magnetic contributions. (b) Illustration of the resulting magnetic structure viewed off one of the edges of the cubic unit cell edges and (c) down the unit cell's diagonal, along the length of the rod. In (b) and (c) the lavender and teal arrows distinguish the two ferromagnetic layers that are coupled antiferromagnetically.}
	\label{fig:neutron}
\end{figure}

Low temperature neutron scattering was collected on the high resolution powder diffractometer, BT-1, at the NIST Center for Neutron Research for \CoV\/ in order to further elucidate its magnetic nature.
Previous neutron diffraction studies on \CoV\/ reported that the magnetic structure consists of ferromagnetic interactions within the layers that couple antiferromagnetically to each other. \cite{Ozerov1972}
At 300\,K, the nuclear structure was refined using the FullProf suite of program and found to agree well with the cubic $Ia\bar{3}d$ structure with Co$^{2+}$ octahedrally coordinated on the 16$a$ site.
Representational analysis was used to fit the 3\,K data with associated magnetic reflections at: 19.2$^{\circ}$, 33.6$^{\circ}$, 60.2$^{\circ}$ and 67.4$^{\circ}$.
All observed magnetic reflections were indexed using $\mathbf{k}=0$.
Symmetry analysis of the potential magnetic structure was performed using \textsc{SARA}h,~\cite{SARAH} which returned two one-dimensional, one two-dimensional and  two three-dimensional representations within the Little Group $G_k$.
The most reasonable fit to the 3\,K diffraction data, shown in Figure \ref{fig:neutron} (a), was obtained using the first two basis vectors of the seventh representation, $\Gamma_7$ of space group Ia$\bar{3}$d, which are listed in Supporting Information Table II.
The resulting topology of this representation returns a result very similar to what was previously reported, but indicates that the magnetic moments are not aligned along exactly the same direction, but are instead aligned along two different directions as seen in Figure \ref{fig:neutron} (b) and (c).
The absolute orientation of the moments are listed in Supporting Information Table III.

\begin{table}[t]
	\centering
	\caption{Relevant crystallographic parameters for the two Co garnets discussed in the text. All lattice parameters, atomic distances, polyhedra angles were determined from refinements against the neutron diffraction data taken at 50\,K. Distortions in polyhedra were determined by calculating the difference between a refined angle and an ideal symmetric octahedral coordination geometry. Values in parentheses indicate one standard deviation.}
	{\renewcommand{\arraystretch}{1.2}	
		\begin{tabularx}{0.45\textwidth}{|lYY|}
			\hline
			\textbf{Composition}                                    & \textbf{Vanadate } & \textbf{Germanate} \\ 
			\hline
			\textbf{$T_N$}                                          & $\approx$8\,K & $\approx$6\,K \\ 
			\hline
			\textbf{Lattice Parameter (\AA)}                        & 12.430(3) & 12.354(4) \\ 
			\hline
			\textbf{d(O--O)$_{intrachain}$ (\AA)}                   & 2.97 & 2.76 \\ 
			\hline
			\multirow{2}{*}{\textbf{d(O--O)$_{interchain}$ (\AA)}}  & 2.87 & 2.97 \\ 
				 											 	    & 2.68 & 2.68 \\ 
			\hline
			\multirow{2}{*}{\textbf{Octahedral Angle ($^{\circ}$)}} & 90.17 & 93.8\\ 
                                                                    & 89.83 & 86.2 \\ 
			\hline
			\textbf{Octahedral Distortion}                          & 0.19\% & 4.2\% \\ 
			\hline                                         
	\end{tabularx}}

	\label{Crystallography}
\end{table}

Despite the similar ordering temperatures, it is interesting to note that \CoGe\/ and \CoV\/ adopt very distinct magnetic structures where \CoGe\/ contains chains of antiparallel spins along the body diagonals whereas \CoV\/ adopts a more two dimensional orientation with ferromagnetic layers that are coupled antiferromagnetically.~\cite{Neer2017}
Clearly the easy axis of the magnetic moments in \CoV\/ is significantly different from \CoGe\/ as the spins lay canted within the planes and show no clear order along the rods as shown in Figure \ref{fig:neutron} (c).
%
%

As seen in in Table \ref{Crystallography}, the changes to the diamagnetic composition of the structure from \CoGe to \CoV\/ result in a significantly larger unit cell that extends all of the interatomic distances and has an impact on the local polyhedral bond angles as well.
Substituting Na$^+$ for Y$^{3+}$ on the cubic sublattice in \CoV,\/ the sublattice's average radii is expanded, resulting in a ~8\% increase in the through-space distance between neighboring Co$^{2+}$ ions compared to the germanate.
Conversely, the average radius of the tetrahedral sublattice in \CoV\/ contracts by ~9\% with the substitution from Ge$^{4+}$ to V$^{5+}$. 

\begin{figure}[b]
	\includegraphics[width=0.3\textwidth]{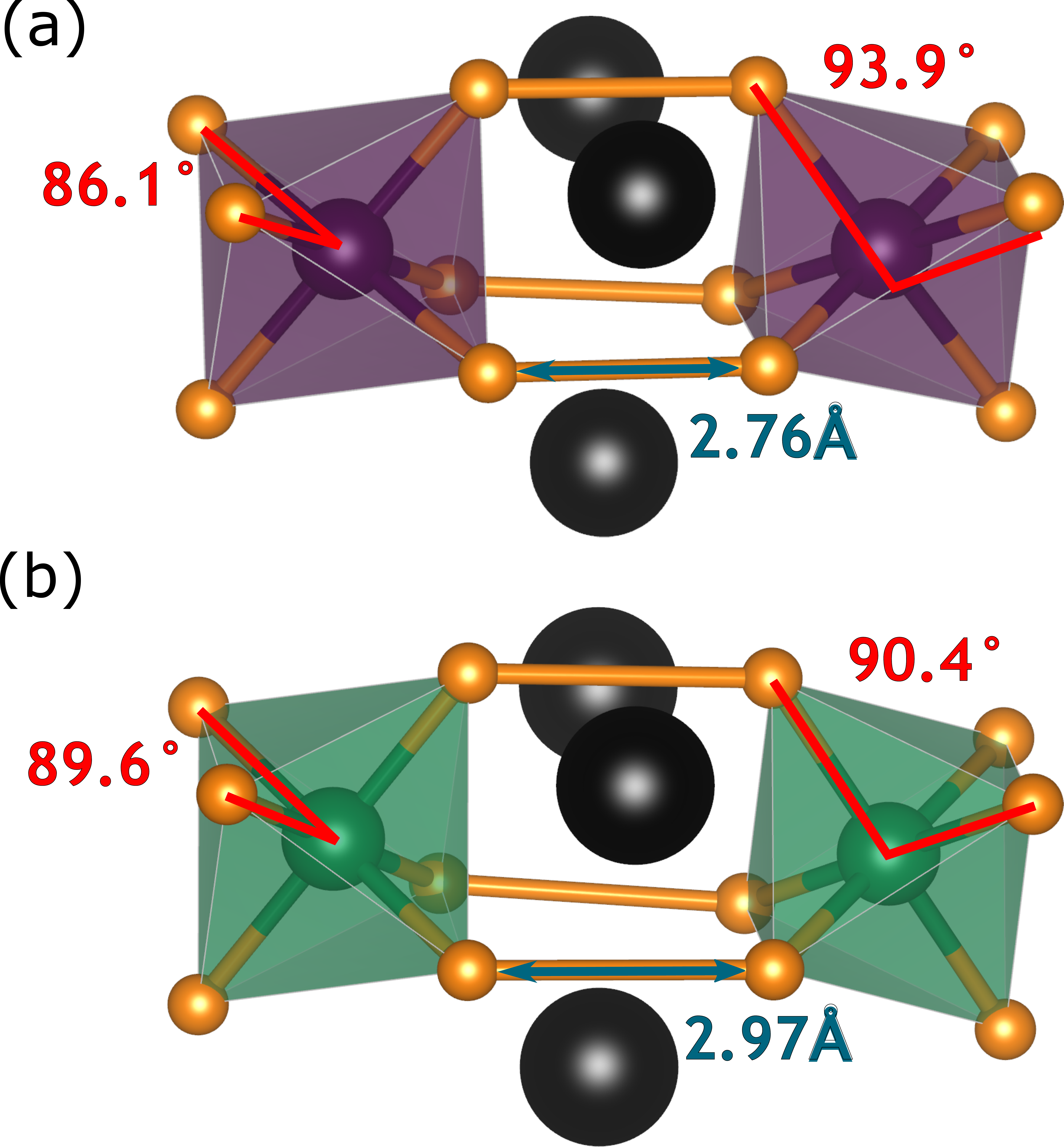}\caption{Angles and distances of nearest neighbors in \CoGe\/ (a) and \CoV\/ (b) garnets along the length of the rods. As the distance between the Co$^{2+}$ decrease the deviations from 90$^{\circ}$ increase, creating a more distorted octahedra in \CoGe, than in \CoV.}
	\label{fig:distort}
\end{figure}

More critically, the octahedra in the vanadate are exhibit significantly less of the trigonal distortion seen in than the germanate, which involves a compression along one edge of the basal plane within the polyhedra, resulting in a less than 0.2\% deviation from 90$^{\circ}$ O--Co--O bond angles as illustrated in Figure \ref{fig:distort}(b).
This change in the local octahedral distortion likely drives the reorientation of the easy axis as the octahedra are no longer compressed along the direction of the rods.

To better understand the impact that these changes to the crystal chemistry have on the electronic structure of the material, density functional theory calculations were performed. 
Figure 7 (a) and (b) show the calculated density of states for CaY$_2$Co$_2$Ge$_3$O$_{12}$ and NaCa$_2$Co$_2$V$_3$O$_{12}$ respectively using the experimentally determined non-collinear magnetic structures. 
In both materials, we find that the top of the valence band consists predominantly hybridized Co $d$ and O $p$ states. 
According to superexchange theory \cite{Harrison2007}, the bandwidth of the noncollinear spin system will be reduced as observed here: The valence bandwidth of NaCa$_2$Co$_2$V$_3$O$_{12}$ is narrower than that of CaY$_2$Co$_2$Ge$_3$O$_{12}$ owing to the non-parallel magnetic configuration in NaCa$_2$Co$_2$V$_3$O$_{12}$.
Despite the total width of the broad valence band, the exchange coupling is determined by (proportional to) the bandwidth of the Co-O hybridized states located near the Fermi level. 
One can find that the width of the two occupied hybridized peaks below $E_f$ are nearly unchanged, indicating that the exchange coupling energies for both systems should be similar to each other.
On the other hand, the conduction band minimum is now dominated by the V $d$ states, resulting in substantial rehybridization and broadening of the conduction band. 
The broader empty states above $E_f$ lead to a stronger coupling in NaCa$_2$Co$_2$V$_3$O$_{12}$ ($\Theta_{CW}$ = -44 K vs -32 K). 

\begin{figure}[b]
	\includegraphics[width=0.4\textwidth]{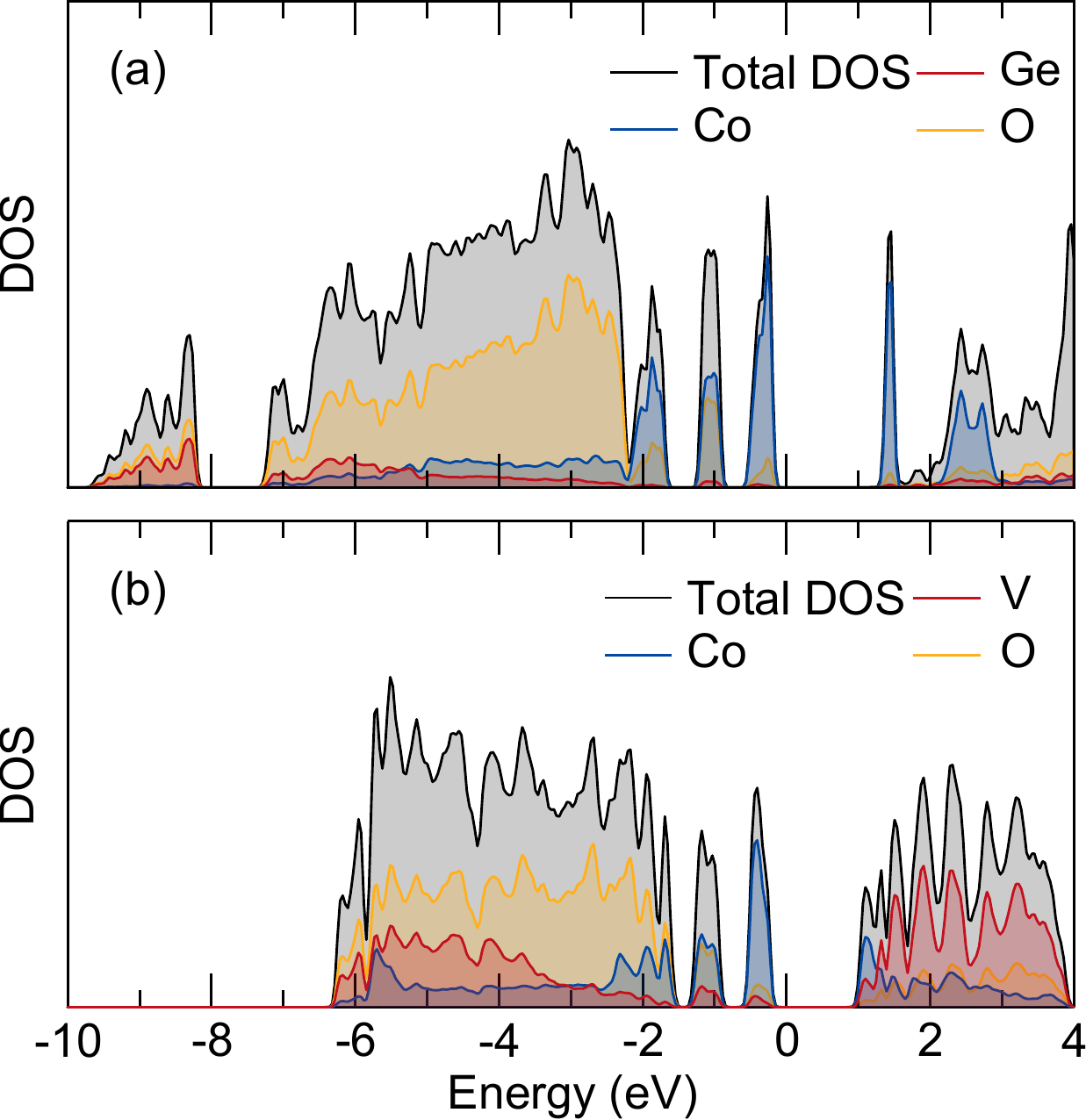}\caption{Calculated densities of state for (a) \CoGe\/ and (b) \CoV.}
	\label{fig:dos}
\end{figure}

\begin{figure} [h!]
	\includegraphics[width=0.4\textwidth]{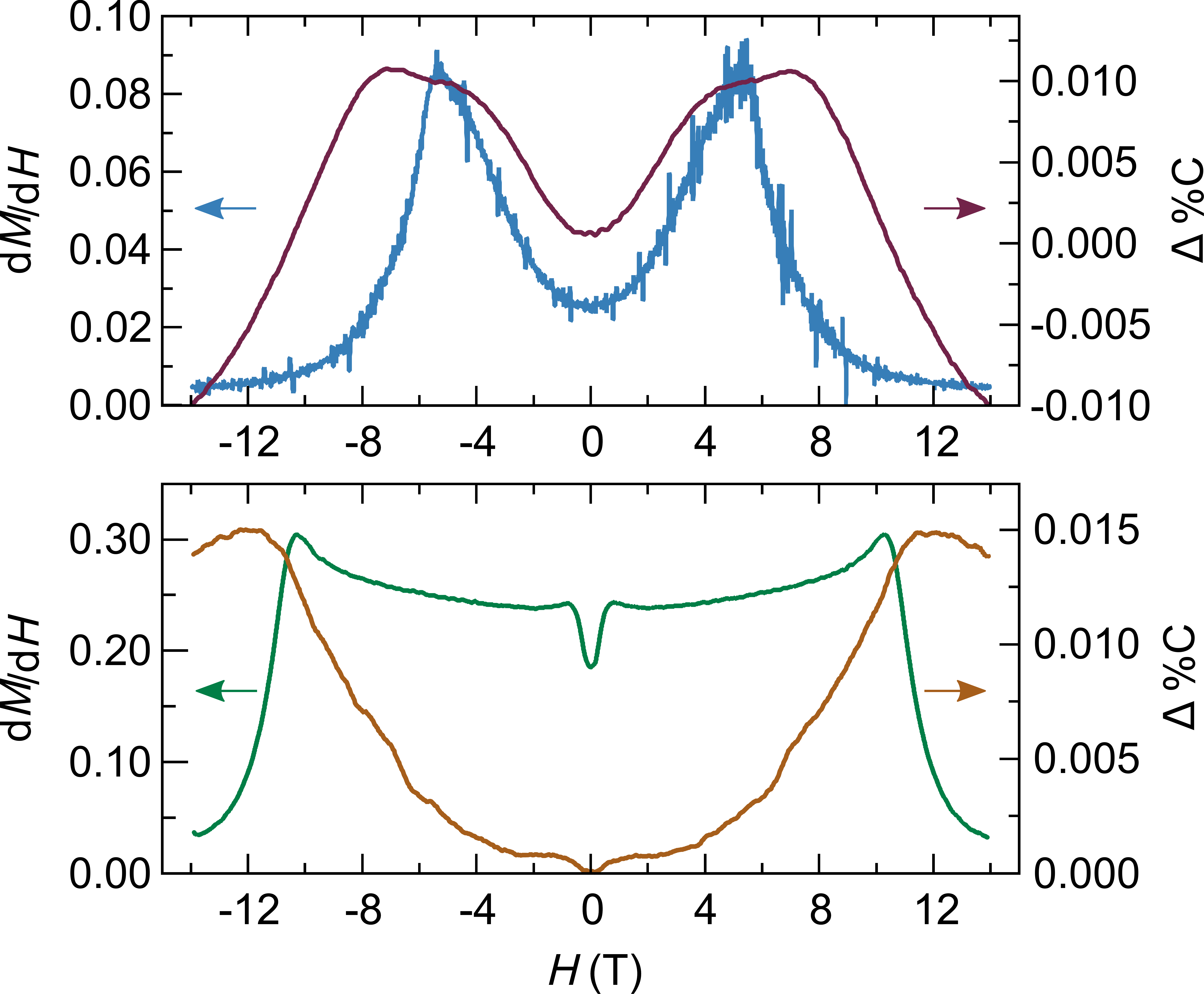}\caption{Field Dependent Capacitance Measurements overlayed on the derivative of the magnetization taken at 2\,K for  (a) CaY$_2$Co$_2$Ge$_3$O$_{12}$ and (b) NaCa$_2$Co$_2$V$_3$O$_{12}$. Note that the features closely correspond to the critical fields indicative of the quantum phase transition for each phase.}\label{fig:map-cap}
\end{figure}

Figure \ref{fig:map-cap} shows that the magnetocapacitance of both phases exhibits a clear dependence on the applied magnetic field.
Given the absence of any structural distortions in the low temperature neutron diffraction, the origin of this dielectric coupling cannot be attributed to a proper ferroelectric coupling due to the cubic symmetry of the unit cell.
For small applied fields both \CoV\/ and \CoGe\/  magnetocapacitance exhibits an approximately quadratic dependence on the magnetic field following the onset of antiferromagnetic order, and this field-dependence quickly decreases in the paramagnetic state (see SI Fig 2 and 3). 
A comparison of the squared magnetization with the magnetocapacitance measurements shows a clear quadratic field dependence between 4\,T in \CoGe, which suggests a coupling of the form $P^2M^2$ that is allowed for all materials regardless of space group symmetry.~\cite{Birol2012}
This type of coupling is a signature of spin-phonon coupling and is reminiscent of what has been reported in CoTiO$_3$,~\cite{Harada2016} NiCr$_2$O$_4$,\cite{Sparks2014} and SeCuO$_3$.\cite{Lawes2003}
More interestingly, significant deviations from this quadratic dependence are found for both compositions near fields associated with the suppression of long-range magnetic order (7\,T and 11\,T for the germanate and vanadate respectively) and produce a change in sign from a positive to negative coupling. 
The most likely explanation for this change in field-dependence is that as the quantum critical transition causes the materials to alter from an antiferromagnetic to a field-polarized paramagnetic state, at which point the coupling becomes more similar to what would be expected of a ferromagnetic material.~\cite{Lawes2009}
These results, taken together, highlight the utility of magnetocapacitance measurements for studying materials with complex magnetic phase diagrams and suggests that other phases exhibiting low-field magnetic transitions should be investigated to better understand this mechanism for magnetodielectric coupling.

\section{Summary}
Here we have studied the nature of the magnetic order in \CoV\/ has been studied using temperature- and field-dependent susceptibility measurements.
The magnetic structure was determined using powder neutron diffraction data, and we find that, compared to \CoGe\/, altering the diamagnetic ions on the cubic and tetrahedral sites results in substantial changes to the local coordination environment of the octahedra in \CoV\/.
These changes directly impact the single ion anisotropy of the Co-ions and results in a reorientation of the easy axis for the ordered moment. 
Whereas \CoGe\/ exhibits order within the rods, the moments in \CoV\/ prefer to align along the $b$-axis of the unit cell to produce a layered antiferromagnetic order. 
Additionally, both \CoV\/ and \CoGe\/ were found to exhibit Quantum Critical Phase transitions that correlate strongly with features found in the magnetocapacitance. 
\section{Acknowledgments}

AJN, VAF, MZ, and BCM gratefully acknowledge support from the Office of Naval Research Grant \#\,N00014-15-1-2411.
M.G. was supported by the U.S. DOE, Office of Basic Energy Sciences, under grant no. DE-SC0012375 and J.M.R. was supported by the National Science Foundation (grant no. DMR-1454688)
We acknowledge the support of the National Institute of Standards and Technology, U. S. Department of Commerce, in providing the neutron research facilities used in this work.
Certain commercial equipment, instruments, or materials are identified in this document. 
Such identification does not imply recommendation or endorsement by the National Institute of Standards and Technology nor does it imply that the products identified are necessarily the best available for the purpose.

\bibliography{SodiumCobalt}
\clearpage
\end{document}